%% file: p400d_sdss_cv_4asp_v03.tex
\documentclass[11pt,twoside]{article}
\usepackage{asp2014}

\def\journal#1#2#3#4{{, #1, { #2}, #3 (#4)}}

\def\astl{Astronomy Letters}

\def\apjl{Astrophys.\ J.\ (Letters)}
\def\apjs{ApJS}
\def\aap{A\&A}

\def\aj{Astron.\ J.}
\def\mnras{MNRAS}

\def\etal{\mbox{et al.}}

\aspSuppressVolSlug
\resetcounters

\begin{document}

\title{Sample of Cataclysmic Variables from 400d X-ray Survey}
\author{R.~Burenin$^1$, \fbox{M.~Revnivtsev$^1$}, A.~Tkachenko$^1$,
  V.~Vorobyev$^1$, A.~Semena$^1$, A.~Meshcheryakov$^1$,
  S.~Dodonov$^2$, M.~Eselevich$^3$, and M.~Pavlinsky$^1$
  \affil{$^1$Space Research Institute, Moscow, 117997 Russia;}
  \affil{$^2$Special Astrophysical Observatory, Nizhnii Arkhyz, 369167
    Russia;} \affil{$^3$Institute of Solar-Terrestrial Physics,
    Irkutsk, 664033 Russia;}}

% This section is for ADS Processing.  There must be one line per author.
\paperauthor{R.~Burenin}{rodion@hea.iki.rssi.ru}{}{Space Research Institute}{}{Moscow}{}{117997}{Russia}
\paperauthor{M.~Revnivtsev}{}{}{Space Research Institute}{}{Moscow}{}{117997}{Russia}
\paperauthor{A.~Tkachenko}{}{}{Space Research Institute}{}{Moscow}{}{117997}{Russia}
\paperauthor{V.~Vorobyev}{}{}{Space Research Institute}{}{Moscow}{}{117997}{Russia}
\paperauthor{A.~Semena}{}{}{Space Research Institute}{}{Moscow}{}{117997}{Russia}
\paperauthor{A.~Meshcheryakov}{}{}{Space Research Institute}{}{Moscow}{}{117997}{Russia}
\paperauthor{S.~Dodonov}{}{}{Institute of Solar-Terrestrial Physics}{}{Nizhnii Arkhyz}{}{369167}{Russia}
\paperauthor{M.~Eselevich}{}{}{Special Astrophysical Observator}{}{Irkutsk}{}{664033}{Russia}
\paperauthor{M.~Pavlinsky}{}{}{Space Research Institute}{}{Moscow}{}{117997}{Russia}

\begin{abstract}
  
  We present a sample of eight cataclysmic variables (CVs) identified
  among the X-ray sources of the 400 square degree (\emph{400d}) X-ray
  ROSAT/PSPC survey. Based on this sample, we have obtained
  preliminary constraints on the X-ray luminosity function of CVs in
  the solar neighbourhood in the range of low luminosities,
  $L_X \sim 10^{29}$--$10^{30}$~erg\,s$^{-1}$ ($0.5$--$2$~keV). We
  show that the logarithmic slope of the CV luminosity function in
  this luminosity range is less steep than that at
  $L_X >10^{31}$~erg\,s$^{-1}$. Our results show that of order of
  thousand CVs will be detected in the SRG/eROSITA all-sky survey at
  high Galactic latitudes, which will allow to obtain much more
  accurate measurements of their X-ray luminosity function.

\end{abstract}

\section{Introduction}

One of the most natural way to select statistically complete samples
of cataclysmic variables (CV) is to detect them as X-ray sources in
large area surveys. Almost all of the objects detected in various
X-ray sky surveys to date \citep[see,
e.g.,][]{sazonov06,pretorius07b,revnivtsev08b} had an X-ray luminosity
$L_X>10^{30}$~erg\,s$^{-1}$ in the 0.5--2~keV energy band.
Measurements of the CV number density at luminosities
$L_X<10^{30}$~erg\,s$^{-1}$ will provide new valuable information for
CV population studies.

To study the X-ray luminosity function of CVs at lower X-ray
luminosities we used the 400 square degree (400d) survey based on
ROSAT/PSPC pointings \citep{400d}. The preliminary results of the
search for CVs among the X-ray sources of this survey were presented
by \cite{ayut15}. In this contribution we discuss the results of CV
X-ray luminosity function measurements based on X-ray CV sample
obtained to date, further details can be found in \cite{br16}.

\section{CV Selection}

To search for point sources we used the central part of the ROSAT
field of view with diameter $<37\arcmin$ in which the angular
resolution is $\leq 70\arcsec$ (FWHM). The average positional accuracy
for an X-ray source is at least 10\arcsec\ (at 95\% confidence). We
used 1605 ROSAT pointings in the 0.5--2~keV band at high Galactic
latitudes $|b|>25^\circ$.

The geometric area of the 400d survey for point sources is 436.7 sq.\
deg. The geometric area of the 400d survey overlapping with SDSS
photometric fields is 262.3 sq.\ deg. At X-ray flux
$\approx 2.5\times10^{-14}$~erg\,s$^{-1}$cm$^{-2}$ (0.5--2~kwV) the
survey area is equal to half of the geometric area. In this survey
more than 37\,000 X-ray sources with fluxes above
$10^{-14}$~erg\,s$^{-1}$cm$^{-2}$ were detected. About 22\,000 of
these sources were detected in the fields where there is an overlap
with SDSS fields.

For the identification of X-ray sources from the 400d survey we used
data from the 12th release of the SDSS \citep{sdss12} and the WISE
infrared all-sky survey \citep{wright10}. The CV optical spectrum
should be dominated in its the blue part by the emission from the WD
and the accretion disk. Therefore, for the selection of CV candidates,
we used the criterion $u^\prime-g^\prime<0.7$. To eliminate a large
number of quasars, which strongly contaminate the CV sample, we
additionally eliminated the objects with colors $w_1-w_2>0.6$, where
$w_1$ and $w_2$ are the photometric bands of the WISE infrared all-sky
survey \citep{wright10}. This spectral range corresponds to the
Rayleigh-Jeans part of the spectrum for the majority of stars,
therefore the CV color shgould be $w_1-w_2\approx 0$.

% This criterion is fairly conservative. It must allow the possible
% contribution of the companion to be taken into account even if it is
% an \emph{L}-type dwarf \citep[see, e.g.,][]{schmidt15}.

We selected CV candidates with $g^\prime$ magnitudes $<20.0$. Even
when the contribution of the accretion disk is small and the WD
contribution dominates in the spectra of CVs, their absolute
magnitudes are $M_{g^\prime}\approx 12$
\citep{gansicke09,mikej14}. Therefore, the limit $g^\prime<20.0$ for
such systems corresponds to a distance of about $400$~pc, which is
approximately equal to the thickness of the Galactic disk.

There are 53 objects satisfying the criteria listed above in the 400d
survey. Some of them(four objects) turned out to be previously known
CVs. To identify the nature of the remaining objects, we carried out
additional optical observations.

\begin{table}[t]
\caption{Cataclysmic variables detected in the 400d survey}
\begin{center}
{\footnotesize
\begin{tabular}{lccclcl}  % l = left, c = centered
\tableline
\noalign{\smallskip}
Name & $m_{g^\prime}$ & $f_X$ & $L_X$ & Alternative name\\
 &  & ~erg\,s$^{-1}$cm$^{-2}$ & ~erg\,s$^{-1}$ & \\

\noalign{\smallskip}
\tableline
\noalign{\smallskip}
\input{cv_table.tex}
\noalign{\smallskip}
\tableline\
\end{tabular}
}
\end{center}
\vskip -16pt
\footnotesize
$^*$ The rough estimate of the X-ray luminosity $L_X$ (for more details, see the text).
\normalsize
\end{table}

\articlefigure[width=.5\textwidth]{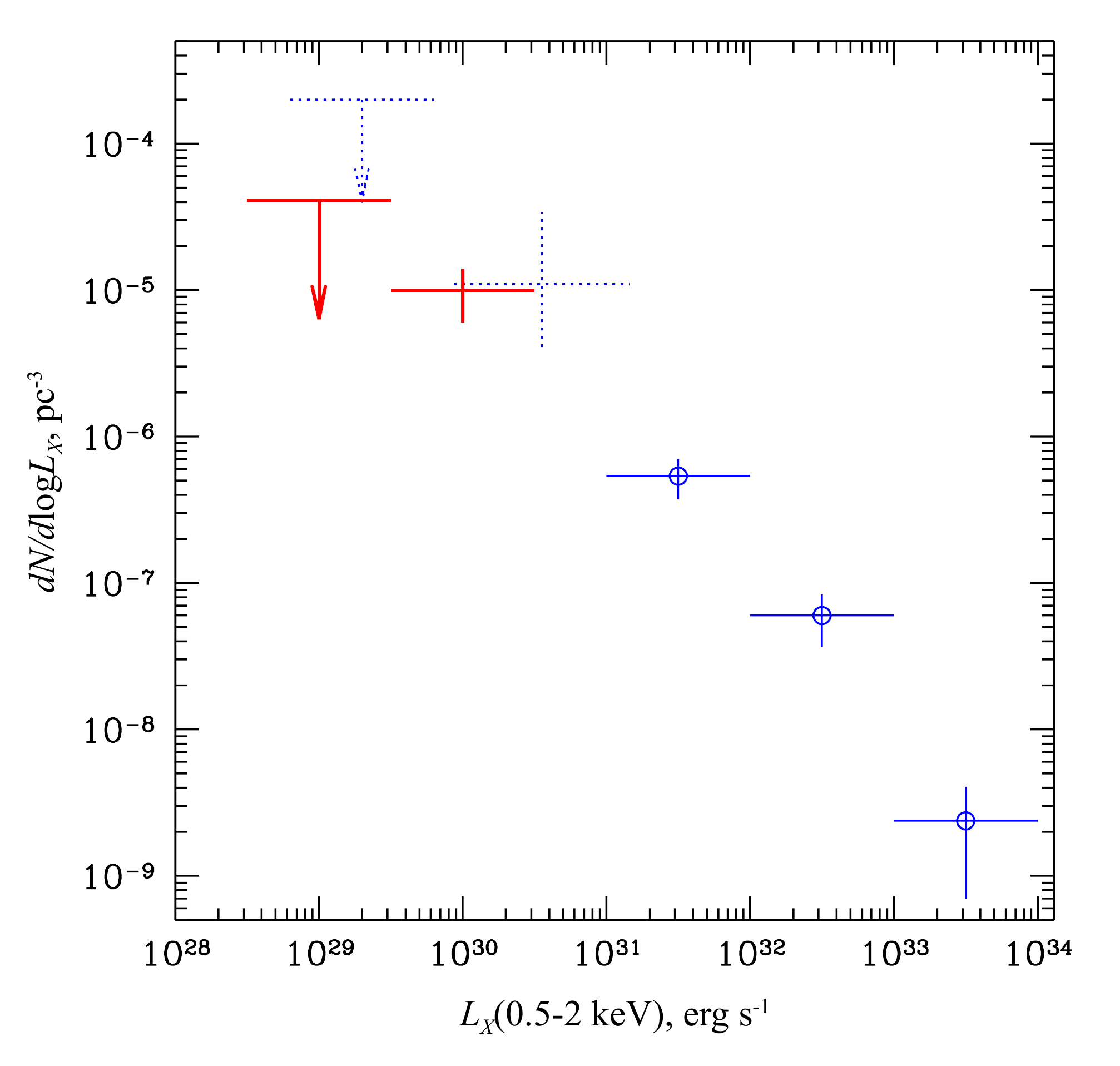}{lf}{Constraints
  on the CV luminosity function. The blue circles indicate the
  measurements of the X-ray luminosity function based on RXTE data
  \citep{sazonov06} recalculated to the 0.5--2~keV energy band under
  the assumption of a power-lawspectrum $dN/dE\propto E^{-1.6}$. The
  dotted blue crosses indicate the constraints on the luminosity
  function from the the ROSAT North Ecliptic Pole survey
  \citep{pretorius07b}.}

\section{Optical Observations of CV Candidates}

The additional optical observations were carried out with the
Russian-Turkish 1.5-m telescope (RTT150) using TFOSC
instrument\footnote{http://hea.iki.rssi.ru/rtt150/en/index.php?page=tfosc},
the 6-m (BTA) telescope at the Special Astrophysical Observatory of
the Russian Academy of Sciences using SCORPIO spectrograph
\citep{scorpio05}, and with the 1.6-m AZT-33IK telescope at the Sayan
Observatory, Russia, using newly installed ADAM spectrograph
\citep{adam16,azt33ik16}. We observed 34 objects, four of which turned
out to be new, previously unknown CVs. The spectra for two of them
were later also obtained in the SDSS. The remaining objects turned out
to be quasars at various redshifts, typically at $z=1$--$2$. Basic
information about the CVs in our sample is presented in the table. To
date we have failed to measure the spectra for 15 objects with
magnitudes $19.5<g^\prime<20$, this work will be continued in future.

\section{Constraints on the CV Luminosity Function in the Solar Neighbourhood}

We detected eight CVs, with more or less reliable distance estimates
being available only for two of them. To make rough, preliminary
estimates of the luminosity function, we assume that the remaining
systems have low X-ray luminosity. In this case, their absolute
magnitude should be near a minimum CV absolute value of
$M_{g^\prime}\approx12$, which is observed due to the presence of a WD
in the system \citep{mikej14}. The X-ray luminosities that are derived
for systems with unknown distances under the assumption of
$M_{g^\prime}\approx12$ are given in the table and marked by an
asterisk. The distance estimates for these systems turn out to be
within the range 200--300~pc. Some of the systems (for example,
400d~j$124325.7\!+\!025541$, 400d~j$152212.8\!+\!080338$ and
400d~j$204720.3\!+\!000008$) can actually have a higher
luminosity. However, the luminosity of all these systems, on average,
cannot be much higher, because, in this case, the objects should be
located at larger distances, but at high Galactic latitudes the space
density of sources drops rapidly at distances greater than the
thickness of the Galactic disk.

Based on these data, we can obtain a preliminary constraint on the
X-ray luminosity function of CVs. 
%For simplicity, we assume that the
%flux limit for X-ray sources in the 400d survey is
%$2.5\times10^{-14}$~erg\,s$^{-1}$cm$^{-2}$, and that the area
%overlapping with the SDSS is 262.3~sq.\ deg. 
The dependence of the density of sources in the Galactic disk on the
perpendicular distance from the Galactic plane,
$\rho (z) = \rho_0 e^{-|z|/h}$, can be taken into account by
calculating a generalized volume \cite{pretorius07b}.  We assume that
the exponential scale height of the Galactic disk for our CVs is
260~pc, which corresponds to the disk scale height for old,
short-period systems \citep{pretorius07a}. The sample incompleteness
can be taken into account statistically by appropriately increasing
the measurement errors.

The X-ray luminosities of the six systems from our sample are
$3\times 10^{29}$--$3\times10^{30}$~erg\,s$^{-1}$. Taking into account
all of the aforesaid, we obtain the following constraint on the CV
space density based on this sample:
$\rho_0 = 1.0\pm 0.4 \times 10^{-5}$~pc$^{-3}$. Only one system,
400d~j$154730.1\!+\!071151$, with a luminosity close to the upper
boundary of the luminosity range
$3\times 10^{28}$--$3\times10^{29}$~erg\,s$^{-1}$ falls within this
range. In this case, the upper limit on the CV space density for this
range is $\rho_0 < 4.1 \times 10^{-5}$~pc$^{-3}$ (at 95\%
confidence). This system may be classified as a pre-cataclysmic
variable. If such systems are not considered as CVs, then the limit on
the CV space density is $\rho_0 < 2.7 \times 10^{-5}$~pc$^{-3}$.

Our constraints on the X-ray luminosity function of CVs are shown by
the solid red crosses in Fig.~\ref{lf}. The constraints from previuos
studies are also shown in this Figure. Our constraint on the CV space
density near $L_X\approx 10^{30}$~erg\,s$^{-1}$ agrees well with that
from \cite{pretorius07b}. We obtained a stronger upper limit on the CV
space density near an X-ray luminosity
$L_X\approx 10^{29}$~erg\,s$^{-1}$. The comparison of our constraints
with the luminosity function at $L_X> 10^{31}$~erg\,s$^{-1}$ derived
previously \citep{sazonov06} shows that its slope becomes less steep
at low luminosities (see Fig.~\ref{lf}).

% Our constraints can be significantly improved if more reliable
% distance estimates will be obtained for a larger number of systems and
% if the statistically complete sample will be expanded to include
% system with a lower optical brightness. 

The area of 400d X-ray survey is approximately 1\% of all-sky, and its
depth is approximately equal to the depth of future SRG/eROSITA
survey. Therefore, our results show that in SRG/eROSITA all-sky survey
of order of thousand CVs will be detected at high Galactic latitudes,
which will allow to obtain much more accurate measurements of their
X-ray luminosity function.

\acknowledgements This work was supported by RSF grant no.\ 14-22-00271.

%\bibliography{editor}  % For BibTex

% For non-BibTex:

\end{document}

%% file: cv_table.tex
400d~j$001912.9\!+\!220736$ & $19.61$ & $4.84\times 10^{-14}$ & $^*6.4\times 10^{29}$ & SDSS J001912.58+220733.0 \\ 
400d~j$050146.2\!-\!035914$ & $18.41$ & $1.78\times 10^{-13}$ & $5.8\times 10^{30}$ & HY Eri                             \\ 
400d~j$124325.7\!+\!025541$ & $17.72$ & $5.51\times 10^{-13}$ & $^*1.3\times 10^{30}$ & 1E 1240.8+0312 \\ 
400d~j$152212.8\!+\!080338$ & $18.99$ & $1.58\times 10^{-13}$ & $^*1.2\times 10^{30}$ & SDSS J152212.20+080340.9  \\ 
400d~j$154730.1\!+\!071151$ & $16.34$ & $1.16\times 10^{-13}$ & $2.0\times 10^{29}$ &  \\ 
400d~j$160002.4\!+\!331120$ & $19.89$ & $8.87\times 10^{-14}$ & $^*1.5\times 10^{30}$ & VW CrB \\ 
400d~j$160547.5\!+\!240524$ & $19.78$ & $5.47\times 10^{-14}$ & $^*8.5\times 10^{29}$ &  \\ 
400d~j$204720.3\!+\!000008$ & $19.36$ & $4.19\times 10^{-14}$ & $^*4.4\times 10^{29}$ & SDSS J204720.76+000007.7  \\ 